\documentclass[%
reprint,
showpacs,preprintnumbers,
nofootinbib,
amsmath,amssymb,
aps,
prd,
floatfix,
]{revtex4-1}

\usepackage{graphicx}
\usepackage{dcolumn}
\usepackage{bm}

\usepackage{url}

\newcommand{\boss}[2]{\ensuremath{\rlap{\kern-2.5pt\ensuremath{\overset{\scriptscriptstyle(-)}{\phantom{#1}}}}{\ensuremath{{#1}_{#2}}}}}

\begin{document}

\author{Carlo Giunti}
\email{giunti@to.infn.it}
\altaffiliation[also at ]{Department of Theoretical Physics, University of Torino, Italy}
\affiliation{INFN, Sezione di Torino, Via P. Giuria 1, I--10125 Torino, Italy}

\author{Marco Laveder}
\email{laveder@pd.infn.it}
\affiliation{Dipartimento di Fisica ``G. Galilei'', Universit\`a di Padova,
and
INFN, Sezione di Padova,
Via F. Marzolo 8, I--35131 Padova, Italy}

\date{\today}


\preprint{\begin{tabular}{l}
EURONU-WP6-10-25
\\
arXiv:1010.1395
\end{tabular}}

\title{Short-Baseline $\bar\nu_{\mu}\to\bar\nu_{e}$ Oscillations}

\begin{abstract}
We analyze the recent results of the MiniBooNE short-baseline experiment
on $\bar\nu_{\mu}\to\bar\nu_{e}$ oscillations in a minimal
model-independent framework of antineutrino mixing
in conjunction with the positive LSND signal and the negative KARMEN measurements.
We show that the data of the three short-baseline
$\bar\nu_{\mu}\to\bar\nu_{e}$
experiments are compatible.
Taking into account also the model-independent constraints
due to the limits on short-baseline
$\bar\nu_{e}$
disappearance obtained in reactor antineutrino experiments,
we find that the favored region of the effective oscillation parameters
lies within
$2\times10^{-3} \lesssim \sin^22\vartheta \lesssim 5\times10^{-2}$
and
$0.2 \lesssim \Delta{m}^2 \lesssim 2 \, \text{eV}^2$.
\end{abstract}

\maketitle

\section{\label{001}Introduction}

The MiniBooNE collaboration
\cite{1007.1150}
recently reported the observation of a signal
of short-baseline $\bar\nu_{\mu}\to\bar\nu_{e}$ transitions
compatible with that observed in the LSND experiment
\cite{hep-ex/0104049}.
The agreement of the MiniBooNE and LSND signals in favor of neutrino oscillations
is remarkable,
because the two experiments observed the signal
of $\bar\nu_{\mu}\to\bar\nu_{e}$ transitions
at different source-detector distances and different neutrino energy ranges.
Since only the ratio of distance and energy is similar in the two experiments
and neutrino oscillations depend just on this ratio
(see Refs.~\cite{hep-ph/9812360,hep-ph/0211462,hep-ph/0310238,hep-ph/0405172,hep-ph/0506083,hep-ph/0606054,GonzalezGarcia:2007ib,Giunti-Kim-2007}),
the neutrino oscillation explanation of the two signals is strongly favored.
On the other hand,
the MiniBooNE collaboration did not observe any signal of short-baseline
$\nu_{\mu}\to\nu_{e}$ transitions
\cite{0812.2243}
compatible with the MiniBooNE and LSND signals of
$\bar\nu_{\mu}\to\bar\nu_{e}$ transitions.
Therefore,
it is possible that the effective parameters which govern neutrino and antineutrino oscillations
are different, maybe because of a violation of the CPT symmetry
\cite{hep-ph/0010178,hep-ph/0108199,hep-ph/0112226,hep-ph/0201080,hep-ph/0201134,hep-ph/0201211,hep-ph/0307127,hep-ph/0308299,hep-ph/0505133,hep-ph/0306226,Laveder:2007zz,0707.4593,0804.2820,0902.1992,0903.4318,0907.5487,0908.2993,1005.4599,1008.4750}.
From a phenomenological point of view, it is interesting to
consider the neutrino and antineutrino sectors independently,
especially in view of possible experimental checks of the
short-baseline $\bar\nu_{\mu}\to\bar\nu_{e}$ signal
\cite{0909.0355,0910.2698,1007.3228,AndreRubbia:NEU2012}.
In this paper we adopt this point of view
and we present the results of a combined fit of the
MiniBooNE and LSND antineutrino data
in favor of
short-baseline $\bar\nu_{\mu}\to\bar\nu_{e}$ transitions,
together with the constraints imposed by the data of the
KARMEN experiment
\cite{hep-ex/0203021}
in which the transitions have not been observed.
We also take into account the model-independent
constraints imposed by the data of reactor
$\bar\nu_{e}$ disappearance experiments.

In the analysis of the data of
$\bar\nu_{\mu}\to\bar\nu_{e}$
oscillation experiments
we consider the simplest case of an effective two-neutrino-like
short-baseline oscillation probability,
similar to that obtained in the case of four-neutrino mixing
(see Refs.~\cite{hep-ph/9812360,hep-ph/0405172,hep-ph/0606054,GonzalezGarcia:2007ib}),
\begin{equation}
P_{\bar\nu_{\mu}\to\bar\nu_{e}}(L/E)
=
\sin^22\vartheta
\sin^2\left(\dfrac{\Delta{m}^2 L}{4 E}\right)
\,,
\label{002}
\end{equation}
where $\Delta{m}^2$ is the relevant neutrino squared-mass difference
and
$\vartheta$ is the effective mixing angle for $\bar\nu_{\mu}\to\bar\nu_{e}$ transitions.

The plan of the paper is as follows.
In Sections~\ref{004} and \ref{006} we present, respectively, the results of the fits of MiniBooNE and LSND antineutrino data,
and in Section~\ref{021} we discuss the results of the combined fit.
In Section~\ref{023} we present the results of the fit of KARMEN data
and in Section~\ref{028} we discuss the results of the combined fit of MiniBooNE, LSND and KARMEN data.
In Section~\ref{030} we discuss the implications of the constraints from reactor $\bar\nu_{e}$ disappearance experiments.
Finally,
in Section~\ref{037} we draw the conclusions.

\begin{table*}[t!]
\begin{center}
\begin{tabular}{ccccccccc}
&
&
MB
&
LS
&
MB+LS
&
KA
&
MB+LS+KA
&
Re
&
(MB+LS+KA)+Re
\\
\hline
 No Osc. & $\chi^{2}$ & $ 21.4 $ & $ 15.0 $ & $ $ & $ 6.8 $ & $ $ & $ 51.0 $ & $ $ \\
 & NDF & $ 16 $ & $ 5 $ & $ $ & $ 8 $ & $ $ & $ 56 $ & $ $ \\
 & GoF & $ 0.16 $ & $ 0.010 $ & $ $ & $ 0.55 $ & $ $ & $ 0.66 $ & $ $ \\
\hline Osc. & $\chi^{2}_{\text{min}}$ & $ 11.7 $ & $ 1.4 $ & $ 14.6 $ & $ 6.4 $ & $ 25.7 $ & $ 48.5 $ & $ 77.3 $ \\
 & NDF & $ 14 $ & $ 2 $ & $ 18 $ & $ 6 $ & $ 26 $ & $ 54 $ & $ 82 $ \\
 & GoF & $ 0.63 $ & $ 0.51 $ & $ 0.69 $ & $ 0.38 $ & $ 0.48 $ & $ 0.69 $ & $ 0.63 $ \\
 & $\sin^22\vartheta_{\text{bf}}$ & $ 0.91 $ & $ 0.0058 $ & $ 0.006 $ & $ 0.0010 $ & $ 1.00 $ & $ 0.042 $ & $ 0.014 $ \\
 & $\Delta{m}^2_{\text{bf}}$ & $ 0.071 $ & $ 8.13 $ & $ 4.57 $ & $ 6.76 $ & $ 0.052 $ & $ 1.86 $ & $ 0.46 $ \\
\hline PG & $\Delta\chi^{2}_{\text{min}}$ & $ $ & $ $ & $ 1.50 $ & $ $ & $ 6.32 $ & $ $ & $ 3.02 $ \\
 & NDF & $ $ & $ $ & $ 2 $ & $ $ & $ 4 $ & $ $ & $ 2 $ \\
 & GoF & $ $ & $ $ & $ 0.47 $ & $ $ & $ 0.18 $ & $ $ & $ 0.22 $ \\
\hline
\end{tabular}
\caption{ \label{003}
Values of
$\chi^{2}$,
number of degrees of freedom (NDF),
goodness-of-fit (GoF)
and
best-fit values
$\sin^22\vartheta_{\text{bf}}$, $\Delta{m}^2_{\text{bf}}$
of the oscillation parameters
obtained from
the fit of various combinations of
MiniBooNE (MB),
LSND (LS),
KARMEN (KA)
and
reactor Bugey and Chooz (Re)
antineutrino data.
The first three lines correspond to the case of no oscillations (No Osc.).
The following five lines correspond to the case $\bar\nu_{\mu}\to\bar\nu_{e}$ oscillations (Osc.).
The last three lines give the parameter goodness-of-fit (PG) \protect\cite{hep-ph/0304176}.
The variations of
$\sin^22\vartheta_{\text{bf}}$ and $\Delta{m}^2_{\text{bf}}$
depending on the fitted data sets
are due to the oscillating character of
$P_{\bar\nu_{\mu}\to\bar\nu_{e}}$
in Eq.~(\ref{002}).
}
\end{center}
\end{table*}

\section{\label{004}MiniBooNE}

\begin{figure}[t!]
\begin{center}
\includegraphics*[bb=5 11 571 571, width=\linewidth]{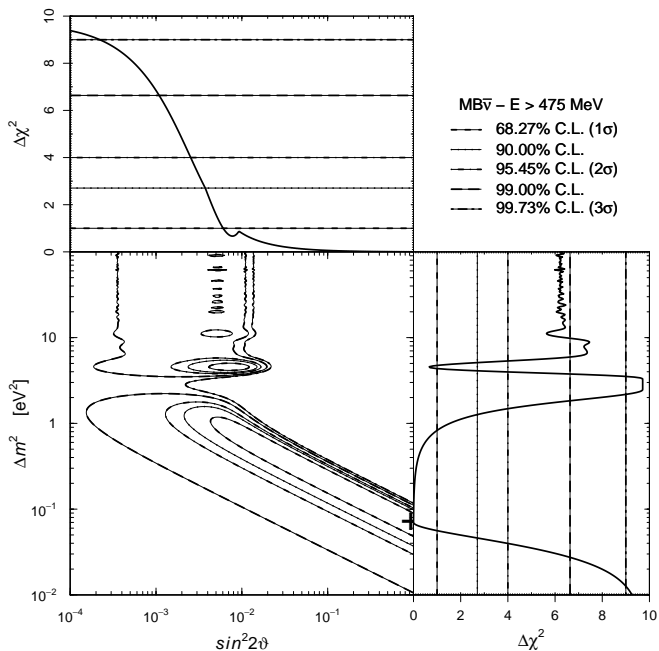}
\end{center}
\caption{ \label{005}
Allowed regions in the
$\sin^{2}2\vartheta$--$\Delta{m}^{2}$ plane
and
marginal $\Delta\chi^{2}=\chi^2-\chi^2_{\text{min}}$'s
for
$\sin^{2}2\vartheta$ and $\Delta{m}^{2}$
obtained from the fit of MiniBooNE (MB) antineutrino data
in the energy range $E>475\,\text{MeV}$ \cite{1007.1150}.
The best-fit point is indicated by a cross.
}
\end{figure}

The MiniBooNE collaboration presented recently the results of a search for
$\bar\nu_{\mu}\to\bar\nu_{e}$
oscillations
obtained with a data sample corresponding to $5.66 \times 10^{20}$ protons on target
\cite{1007.1150}.
The MiniBooNE detector is located at a distance of 541 m from the neutrino source.
The neutrino energy spectrum for the oscillation analysis ranges from 475 MeV to 3 GeV.
Hence, the ratio $L/E$ from which the oscillation probability in Eq.~(\ref{002}) depends
ranges from 0.18 to 1.14 m/MeV,
leading to a sensitivity to $\bar\nu_{\mu}\to\bar\nu_{e}$ transitions
for
$\Delta{m}^2 \gtrsim 10^{-1} \, \text{eV}^2$
appropriate for checking the signal observed in the LSND experiment
\cite{hep-ex/0104049}
(see Section~\ref{006}).

The excess of $\bar\nu_{e}$-like events found by the MiniBooNE collaboration
agrees with the excess found in the LSND experiment
at different source-detector distance and neutrino energy,
but similar ratio $L/E$.
This is a strong indication in favor of the neutrino oscillation explanation
of the two signals,
analogous to the confirmation of solar neutrino oscillations by the
very-long-baseline KamLAND reactor experiment
\cite{0801.4589}
and
the confirmation of atmospheric neutrino oscillations by the
long-baseline K2K \cite{hep-ex/0606032} and MINOS \cite{0806.2237} accelerator experiments.

In this paper we fit the MiniBooNE data reported in Fig.~1 of Ref.~\cite{1007.1150}\footnote{
We would like to thank W.C.~Louis for sending us the precise values of the data in Fig.~1 of Ref.~\cite{1007.1150}.
}
using the method and data given in the MiniBooNE data release in Ref.~\cite{AguilarArevalo:2008rc-dr},
which are relative to the previous MiniBooNE publication \cite{0904.1958}
on the search for $\bar\nu_{\mu}\to\bar\nu_{e}$
(the MiniBooNE data release relative to Ref.~\cite{1007.1150} is still not available).
We rescaled the signal predicted with the method
described in Ref.~\cite{AguilarArevalo:2008rc-dr}
from the $3.39 \times 10^{20}$ protons on target corresponding to the sample in Ref.~\cite{0904.1958}
to the $5.66 \times 10^{20}$ protons on target corresponding to the sample in Ref.~\cite{1007.1150}.
The fractional covariance matrix of systematic uncertainties should be similar in the two data releases.
We corrected the statistical part of the covariance matrix by taking into account the
different number of background events.
In the fit we consider not only the $\bar\nu_{e}$ MiniBooNE data,
but also the $\bar\nu_{\mu}$ data,
which are important because of the correlated uncertainties.
Since the $\bar\nu_{\mu}$ data obtained with $5.66 \times 10^{20}$ protons on target
are not available,
we consider the $\bar\nu_{\mu}$ data
given in the MiniBooNE data release in Ref.~\cite{AguilarArevalo:2008rc-dr}.
The difference is not crucial,
because the $\bar\nu_{\mu}$ data have only an indirect effect on the measurement
of the $\bar\nu_{\mu}\to\bar\nu_{e}$ signal
through the correlated uncertainties.

The results of the least-squares fit of MiniBooNE data
are presented in the first column of Tab.~\ref{003}
and in Fig.~\ref{005}.
The best-fit values of the oscillation parameters and the allowed regions in the
$\sin^22\vartheta$--$\Delta{m}^2$
plane are similar to those obtained by the MiniBooNE collaboration
\cite{1007.1150}.
The goodness-of-fit in the case of no oscillations
may seem too high in comparison with that given in Ref.~\cite{1007.1150}
and not sufficient to require oscillations.
Since in our calculation we fit both the $\bar\nu_{e}$ and $\bar\nu_{\mu}$ data
we have 16 degrees of freedom,
with
$\chi^2 = 21.4$.
However,
most of the $\chi^2$ is due to the $\bar\nu_{e}$ data,
which have only 8 degrees of freedom,
corresponding to the 8 energy bins in Fig.~1 of Ref.~\cite{1007.1150}
for $E>475\,\text{MeV}$.
If we restrict the $\chi^2$ to the six energy bins from
$475$ to $1300\,\text{MeV}$
we have
$\chi^2 = 16.8$,
which is similar to the
$\chi^2 = 18.5$
reported in Ref.~\cite{1007.1150}
for the energy range from $475$ to $1250\,\text{MeV}$.
Therefore,
we agree with Ref.~\cite{1007.1150} on the opinion that a
background-only fit of MiniBooNE data is disfavored.

The first column of Tab.~\ref{003} shows that the oscillation hypothesis
fits the MiniBooNE data with a $\chi^2$ much lower than in the case of no oscillations,
improving significantly the goodness-of-fit.
The decrease of the $\chi^2$ with respect to the case of no oscillations
is mainly due to the improved fit of the six $\bar\nu_{e}$ energy bins from
$475$ to $1300\,\text{MeV}$
which give a contribution to $\chi^2_{\text{min}}$ of
$7.2$,
in approximate agreement with the 8.0
reported in Ref.~\cite{1007.1150}
for the energy range from $475$ to $1250\,\text{MeV}$.

From Fig.~\ref{005} one can see that,
although the best-fit value of $\sin^22\vartheta$ is close to unity,
in practice all the allowed straight region in the log-log plot
ranging
from
$\sin^22\vartheta \approx 1$ and $\Delta{m}^2 \approx 6\times10^{-2}\,\text{eV}^2$
to
$\sin^22\vartheta \approx 2\times10^{-3}$ and $\Delta{m}^2 \approx 1\,\text{eV}^2$,
as well as a small area around
$\sin^22\vartheta \approx 7\times10^{-3}$ and $\Delta{m}^2 \approx 5\,\text{eV}^2$,
are equally plausible.
This is important,
because large values of the effective mixing angle are excluded by
the limits on $\bar\nu_{e}$ disappearance obtained in reactor antineutrino experiments,
as explained in Section~\ref{030}.

\section{\label{006}LSND}

\begin{figure}[t!]
\begin{center}
\includegraphics*[bb=5 11 571 571, width=\linewidth]{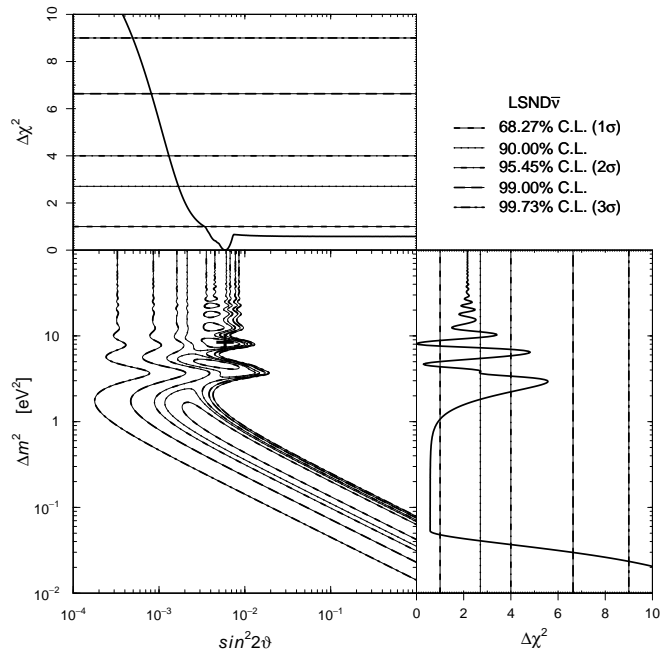}
\end{center}
\caption{ \label{007}
Allowed regions in the
$\sin^{2}2\vartheta$--$\Delta{m}^{2}$ plane
and
marginal $\Delta\chi^{2}$'s
for
$\sin^{2}2\vartheta$ and $\Delta{m}^{2}$
obtained from the fit of LSND antineutrino data.
The best-fit point is indicated by a cross.
}
\end{figure}

The LSND experiment
\cite{hep-ex/0104049}
observed an excess of $\bar\nu_{e}$ events coming from possible
$\bar\nu_{\mu}\to\bar\nu_{e}$ transitions
in a beam of $\bar\nu_{\mu}$ produced by $\mu^{+}$
decay at rest,
\begin{equation}
\mu^{+} \to e^{+} + \nu_{e} + \bar\nu_{\mu}
\,.
\label{008}
\end{equation}
The energy spectrum of $\bar\nu_{\mu}$ is given by
(see Ref.~\cite{Fukugita:2003en})
\begin{equation}
\phi_{\bar\nu_{\mu}}(E)
\propto
E^2 \left( 3 - 4 E / m_{\mu} \right)
\,,
\label{009}
\end{equation}
for $E$ smaller than
\begin{equation}
E_{\text{max}} = (m_{\mu}-m_{e})/2 \simeq 52.6 \text{MeV}
\,.
\label{010}
\end{equation}

The $\bar\nu_{e}$'s
produced by $\bar\nu_{\mu}\to\bar\nu_{e}$ transitions
were detected at an average distance of 30 m through the inverse neutron decay process
\begin{equation}
\bar\nu_{e} + p \to n + e^{+}
\label{011}
\end{equation}
in a detector consisting of an approximately cylindrical tank
8.3 m long by 5.7 m in diameter,
filled with liquid scintillator.
Hence, the oscillation distance varied
between
$L_{\text{min}}=25.85\,\text{m}$
and
$L_{\text{max}}=34.15\,\text{m}$.
The cross section of the detection process in Eq.~(\ref{011}) is
(see Refs.~\cite{hep-ph/0107277,Fukugita:2003en,Giunti-Kim-2007})
\begin{equation}
\sigma_{\bar\nu_{e}p}(E_{e})
\propto
E_{e} p_{e}
\,,
\label{012}
\end{equation}
where $E_{e}$ and $p_{e}$ are, respectively, the positron energy and momentum.
Neglecting the small recoil energy of the neutron,
the positron energy $E_{e}$ is related to the neutrino energy $E$ by
\begin{equation}
E_{e} = E + m_{p} - m_{n} = E - 1.293 \, \text{MeV}
\,,
\label{013}
\end{equation}
where $m_{p}$ and $m_{n}$ are, respectively, the proton and neutron masses.
The neutrino energy threshold is given by
\begin{equation}
E_{\text{min}}
=
\frac{ \left( m_{n} + m_{e} \right)^{2} - m_{p}^{2} }{ 2 \, m_{p} }
=
1.806 \, \text{MeV}
\,.
\label{014}
\end{equation}

The LSND detector had a positron energy resolution which we assume to be Gaussian:
\begin{equation}
R(E_{e}',E_{e})
=
\dfrac{1}{\sqrt{2\pi}\delta_{E_{e}}}
\,
\exp\left(
-
\dfrac{\left(E_{e}'-E_{e}\right)^2}{2\delta_{E_{e}}^2}
\right)
\,,
\label{015}
\end{equation}
with \cite{hep-ex/0104049}
\begin{equation}
\delta_{E_{e}}
=
3.3 \, \text{MeV}
\sqrt{\dfrac{E_{e}}{50\,\text{MeV}}}
\,.
\label{016}
\end{equation}

We fit the LSND data in Fig.~16 of Ref.~\cite{hep-ex/0104049},
which gives the measured $\bar\nu_{e}$ events
$N_{\bar\nu_{e}}^{i}$
in five bins of measured positron energy,
their uncertainties
$\delta{N}_{\bar\nu_{e}}^{i}$
and the expected number of background events
$B_{\bar\nu_{e}}^{i}$.
The fit is obtained by minimizing the least-square function
\begin{equation}
\chi^2
=
\sum_{i=1}^{5}
\left(
\dfrac
{B_{\bar\nu_{e}}^{i} + \eta N_{\bar\nu_{\mu}\to\bar\nu_{e}}^{i} - N_{\bar\nu_{e}}^{i}}
{\delta{N}_{\bar\nu_{e}}^{i}}
\right)^2
+
\left(
\dfrac{\eta-1}{\delta\eta}
\right)^2
\,.
\label{017}
\end{equation}
The number $N_{\bar\nu_{\mu}\to\bar\nu_{e}}^{i}$ of
$\bar\nu_{\mu}\to\bar\nu_{e}$
events in the $i$th bin of measured positron energy
between
$E_{e}^{i} - \Delta{E}_{e}/2$
and
$E_{e}^{i} + \Delta{E}_{e}/2$
is given by
\begin{widetext}
\begin{equation}
N_{\bar\nu_{\mu}\to\bar\nu_{e}}^{i}
=
N_{\bar\nu_{\mu}\to\bar\nu_{e}}^{0}
\frac{
\int_{L_{\text{min}}}^{L_{\text{max}}} \text{d}L \,
L^{-2}
\int_{E_{\text{min}}}^{E_{\text{max}}} \text{d}E \,
\phi_{\bar\nu_{\mu}}(E)
P_{\bar\nu_{\mu}\to\bar\nu_{e}}(L/E)
\sigma_{\bar\nu_{e}p}(E_{e}(E))
\int_{E_{e}^{i} - \Delta{E}_{e}/2}^{E_{e}^{i} + \Delta{E}_{e}/2} \text{d}E_{e}' \,
R(E_{e}',E_{e}(E))
}{
\left(
\frac{1}{L_{\text{min}}}
-
\frac{1}{L_{\text{max}}}
\right)
\int_{E_{\text{min}}}^{E_{\text{max}}} \text{d}E \,
\phi_{\bar\nu_{\mu}}(E)
\sigma_{\bar\nu_{e}p}(E_{e}(E))
\int_{E_{e}^{\text{min}}}^{E_{e}^{\text{max}}} \text{d}E_{e}' \,
R(E_{e}',E_{e}(E))
}
\,,
\label{018}
\end{equation}
\end{widetext}
where
\begin{equation}
N_{\bar\nu_{\mu}\to\bar\nu_{e}}^{0}
=
1.29 \times 10^{4}
\label{019}
\end{equation}
is the number of events expected for 100\% $\bar\nu_{\mu}\to\bar\nu_{e}$ transmutation
in the measured energy range between
$E_{e}^{\text{min}}=20\,\text{MeV}$
and
$E_{e}^{\text{max}}=60\,\text{MeV}$.
We obtained this number by dividing
the $\bar\nu_{\mu}\to\bar\nu_{e}$ excess in Fig.~16 of Ref.~\cite{hep-ex/0104049}
(34.1 events)
by the probability in Tab.~XI of Ref.~\cite{hep-ex/0104049}
($2.64\times10^{-3}$).
The factor $\eta$ in Eq.~(\ref{017})
is introduced in order to take into account the relative uncertainty of
$N_{\bar\nu_{\mu}\to\bar\nu_{e}}^{0}$ \cite{hep-ex/0104049}:
\begin{equation}
\delta\eta
=
0.1
\,.
\label{020}
\end{equation}

Figure~\ref{007} shows the allowed regions in the
$\sin^22\vartheta$--$\Delta{m}^2$
plane
that we obtained from the minimization of $\chi^2$,
with the best-fit values of the oscillation parameters
in the second column of Tab.~\ref{003}.
The allowed regions in Fig.~\ref{007}
are similar to those presented by the LSND collaboration
in Ref.~\cite{hep-ex/0104049},
with some differences due to the fact that we fitted a data set which is smaller than that
used by the LSND collaboration.
Similar problems have been encountered in the fits presented in Refs.~\cite{hep-ex/0203023,hep-ph/0305255}.
In particular,
the allowed straight region in the log-log plot
ranging
from
$\sin^22\vartheta \approx 1$ and $\Delta{m}^2 \approx 5\times10^{-2}\,\text{eV}^2$
to
$\sin^22\vartheta \approx 10^{-3}$ and $\Delta{m}^2 \approx 2\,\text{eV}^2$
and the allowed region at large values of $\Delta{m}^2$
are similar to that obtained by the LSND collaboration
(see Fig.~27 of Ref.~\cite{hep-ex/0104049}).
There is some discrepancy at intermediate values of $\Delta{m}^2$,
where we find two favorite regions at
$\Delta{m}^2 \approx 5\,\text{eV}^2$
and
$\Delta{m}^2 \approx 8\,\text{eV}^2$,
which however are similar to those obtained in Ref.~\cite{hep-ph/0305255}
from a fit similar to ours.
Hence,
we think that
the allowed regions in Fig.~\ref{007} are fairly representative of the
parameter space allowed by the LSND data.

\section{\label{021}Combined Fit of MiniBooNE and LSND Data}

\begin{figure}[t!]
\begin{center}
\includegraphics*[bb=5 11 571 571, width=\linewidth]{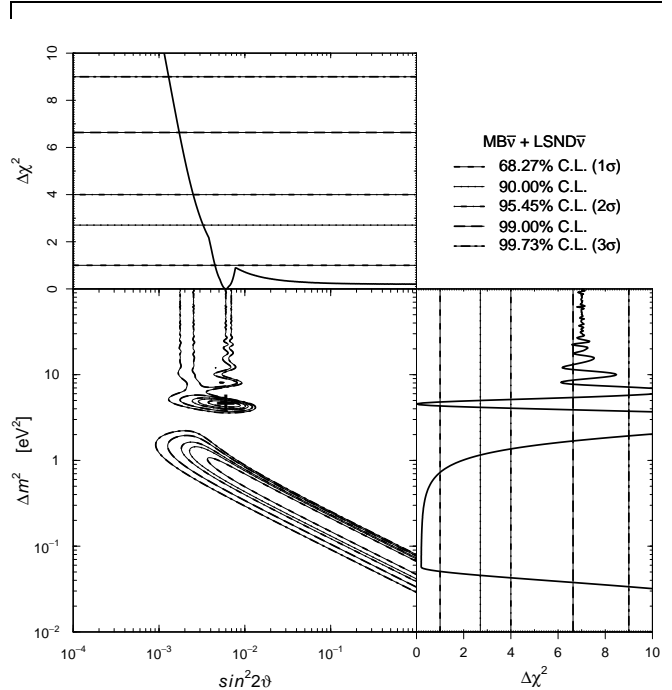}
\end{center}
\caption{ \label{022}
Allowed regions in the
$\sin^{2}2\vartheta$--$\Delta{m}^{2}$ plane
and
marginal $\Delta\chi^{2}$'s
for
$\sin^{2}2\vartheta$ and $\Delta{m}^{2}$
obtained from the combined fit of MiniBooNE (MB) and LSND antineutrino data.
The best-fit point is indicated by a cross.
}
\end{figure}

Comparing Figs.~\ref{005} and \ref{007} one can see that there is a remarkable agreement between
the MiniBooNE and the LSND allowed regions
in the $\sin^22\vartheta$--$\Delta{m}^2$
plane.
The results of the combined fit are given in Fig.~\ref{022} and in the third column of Tab.~\ref{003}.
The excellent parameter goodness-of-fit (PG) \cite{hep-ph/0304176}
of the combined fit quantifies the good compatibility of MiniBooNE and LSND data.
From Fig.~\ref{022}
one can see that the combined fit favors
the allowed straight region in the log-log plot
ranging
from
$\sin^22\vartheta \approx 1$ and $\Delta{m}^2 \approx 5\times10^{-2}\,\text{eV}^2$
to
$\sin^22\vartheta \approx 2\times10^{-3}$ and $\Delta{m}^2 \approx 1\,\text{eV}^2$
and an island at
$\sin^22\vartheta \approx 6\times10^{-3}$ and $\Delta{m}^2 \approx 5\,\text{eV}^2$.

\section{\label{023}KARMEN}

\begin{figure}[t!]
\begin{center}
\includegraphics*[bb=7 14 563 569, width=\linewidth]{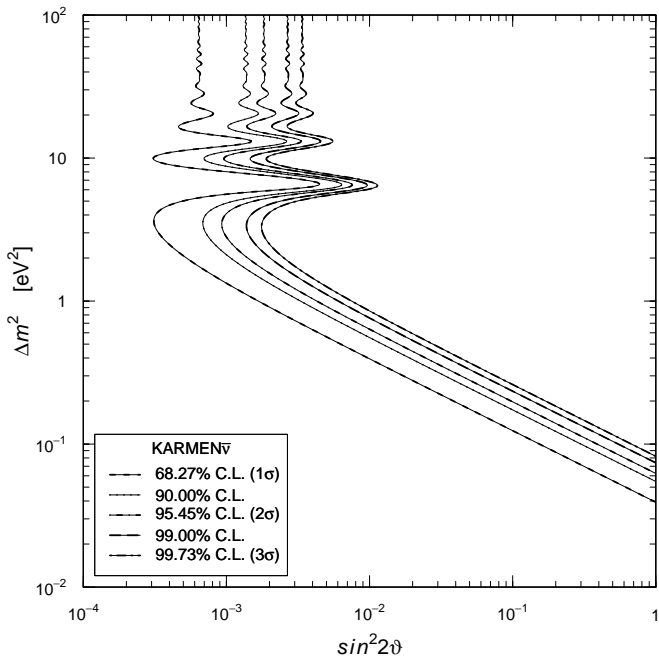}
\end{center}
\caption{ \label{024}
Exclusion curves in the
$\sin^{2}2\vartheta$--$\Delta{m}^{2}$ plane
obtained from the fit of KARMEN antineutrino data.
}
\end{figure}

The KARMEN experiment
\cite{hep-ex/0203021}
searched for $\bar\nu_{\mu}\to\bar\nu_{e}$ transitions
using a beam of $\bar\nu_{\mu}$ produced by the process of $\mu^{+}$
decay at rest in Eq.~(\ref{008})
and detected through the inverse neutron decay process in Eq.~(\ref{011}).
Since these processes are the same as those in the LSND experiments,
the fit of KARMEN is analogous of that described in Section~\ref{006}
for the fit of LSND data.
In the KARMEN experiment the oscillation distance varied
between
$L_{\text{min}}=15.935\,\text{m}$
and
$L_{\text{max}}=19.465\,\text{m}$.
The energy resolution uncertainty was
\begin{equation}
\delta_{E_{e}}
=
0.115 \, \text{MeV}
\sqrt{\dfrac{E_{e}}{\text{MeV}}}
\,.
\label{025}
\end{equation}
We fit the KARMEN data in Fig.~11b of Ref.~\cite{hep-ex/0203021},
which gives the measured $\bar\nu_{e}$ events
$N_{\bar\nu_{e}}^{i}$
in nine bins of measured positron energy
and the expected number of background events
$B_{\bar\nu_{e}}^{i}$.
The number of events expected for 100\% $\bar\nu_{\mu}\to\bar\nu_{e}$ transmutation was
\begin{equation}
N_{\bar\nu_{\mu}\to\bar\nu_{e}}^{0}
=
5826 \left( 1 \pm 0.0923 \right)
\,.
\label{026}
\end{equation}
Since in some energy bins in Fig.~11b of Ref.~\cite{hep-ex/0203021}
the number of measured events is zero,
we perform the fit by minimizing the least-square function
\cite{Baker:1983tu}
\begin{widetext}
\begin{equation}
\chi^2
=
2
\sum_{i=1}^{9}
\left[
B_{\bar\nu_{e}}^{i} + \eta N_{\bar\nu_{\mu}\to\bar\nu_{e}}^{i} - N_{\bar\nu_{e}}^{i}
+
N_{\bar\nu_{e}}^{i}
\ln\left(
\dfrac
{N_{\bar\nu_{e}}^{i}}
{B_{\bar\nu_{e}}^{i} + \eta N_{\bar\nu_{\mu}\to\bar\nu_{e}}^{i}}
\right)
\right]
+
\left(
\dfrac{\eta-1}{\delta\eta}
\right)^2
\,,
\label{027}
\end{equation}
\end{widetext}
with $\delta\eta=0.0923$ from Eq.~(\ref{026}).

Figure~\ref{024} and the fourth column of Tab.~\ref{003} give the result of
the fit of KARMEN data.
The best-fit values of the oscillation parameters
in Tab.~\ref{003}
and the exclusion curves in Fig.~\ref{024}
are in agreement with those found by the KARMEN collaboration
\cite{hep-ex/0203021},
as well as with the results of the fits presented in Refs.~\cite{hep-ex/0203023,hep-ph/0305255}.

\section{\label{028}Combined Fit of MiniBooNE, LSND and KARMEN Data}

\begin{figure}[t!]
\begin{center}
\includegraphics*[bb=5 11 571 571, width=\linewidth]{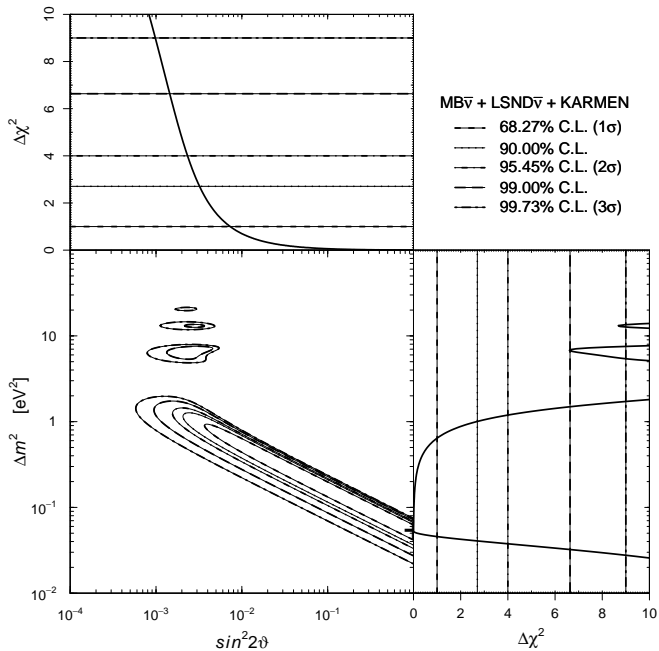}
\end{center}
\caption{ \label{029}
Allowed regions in the
$\sin^{2}2\vartheta$--$\Delta{m}^{2}$ plane
and
marginal $\Delta\chi^{2}$'s
for
$\sin^{2}2\vartheta$ and $\Delta{m}^{2}$
obtained from the combined fit of MiniBooNE (MB), LSND and KARMEN antineutrino data.
The best-fit point is indicated by a cross.
}
\end{figure}

The results of the combined Fit of MiniBooNE, LSND and KARMEN data
on $\bar\nu_{\mu}\to\bar\nu_{e}$ oscillations are given in the fifth column of Tab.~\ref{003}
and in Fig.~\ref{029}.

Comparing Figs.~\ref{022} and \ref{029} one can see that
the inclusion in the analysis of KARMEN data
has mainly the effect of disfavoring the regions at $\Delta{m}^2 \gtrsim 3 \, \text{eV}^2$
allowed by MiniBooNE and LSND data.
The straight region in the log-log plot
ranging
from
$\sin^22\vartheta \approx 1$ and $\Delta{m}^2 \approx 5\times10^{-2}\,\text{eV}^2$
to
$\sin^22\vartheta \approx 10^{-3}$ and $\Delta{m}^2 \approx 2\,\text{eV}^2$
allowed by MiniBooNE and LSND data suffers only a small push towards smaller values of
$\sin^22\vartheta$.
The best-fit point lies in this region, close to the
large-$\sin^22\vartheta$
and
small-$\Delta{m}^2$
edge.
However,
from the marginal
$\Delta\chi^{2}=\chi^2-\chi^2_{\text{min}}$'s
for
$\sin^{2}2\vartheta$ and $\Delta{m}^{2}$
one can see that in practice all the straight region is equally favored.
This is important for the compatibility with the reactor limits on $\sin^{2}2\vartheta$
discussed in the next Section.

\section{\label{030}Constraints from Reactor $\bar\nu_{e}$ Disappearance}

Several reactor experiments have searched for the short-baseline disappearance of $\bar\nu_{e}$'s
(see Refs.~\cite{hep-ph/0107277,Giunti-Kim-2007}),
without positive results
(apart from the hint discussed in Ref.~\cite{0711.4222}).
Such a lack of short-baseline $\bar\nu_{e}$ disappearance
constrains the probability of all transitions of $\bar\nu_{e}$ to other flavor antineutrinos
and all transitions from other flavor antineutrinos to $\bar\nu_{e}$.
We are interested in particular in $\bar\nu_{\mu}\to\bar\nu_{e}$ oscillations,
which are of the second type.
The constraint is model-independent and does not require any assumption
on the type of mixing and on the number of massive neutrinos,
because it follows from simple particle conservation,
which is a characteristic of oscillations.
In fact, since in neutrino oscillations a $\bar\nu_{e}$ must come from an antineutrino of some flavor,
the sum over the probabilities of transition of any flavor antineutrino into $\bar\nu_{e}$ is equal to unity:
\begin{equation}
\sum_{\alpha} P_{\bar\nu_{\alpha}\to\bar\nu_{e}}
=
1
\,.
\label{033}
\end{equation}
Then we have the inequality
\begin{equation}
P_{\bar\nu_{\mu}\to\bar\nu_{e}}
\leq
1 - P_{\bar\nu_{e}\to\bar\nu_{e}}
\,.
\label{034}
\end{equation}
Hence the lower limits obtained in short-baseline reactor antineutrino experiments on
$P_{\bar\nu_{e}\to\bar\nu_{e}}$
imply model-independent upper limits on
$P_{\bar\nu_{\mu}\to\bar\nu_{e}}$.

Considering the simplest case of an effective two-neutrino-like
short-baseline $\bar\nu_{e}$ survival probability
which is governed by the same $\Delta{m}^2$ relevant for
the effective short-baseline probability of
$\bar\nu_{\mu}\to\bar\nu_{e}$
transitions in Eq.~(\ref{002}),
we have
\begin{equation}
P_{\bar\nu_{e}\to\bar\nu_{e}}(L/E)
=
1
-
\sin^22\vartheta_{ee}
\sin^2\left(\dfrac{\Delta{m}^2 L}{4 E}\right)
\,,
\label{035}
\end{equation}
where
$\vartheta_{ee}$ is the effective mixing angle,
which can be different from that of $\bar\nu_{\mu}\to\bar\nu_{e}$ transitions
(which we have denoted for simplicity $\vartheta$,
but could have been called more appropriately
$\vartheta_{e\mu}$).
In this case,
the inequality in Eq.~(\ref{034}) implies
\begin{equation}
\sin^22\vartheta \leq \sin^22\vartheta_{ee}
\,.
\label{036}
\end{equation}
Therefore, the exclusion curves obtained in short-baseline reactor antineutrino experiments
which place upper limits on the value $\sin^22\vartheta_{ee}$
as a function of $\Delta{m}^2$ imply model-independent upper limits on the value of $\sin^22\vartheta$
in short-baseline $\bar\nu_{\mu}\to\bar\nu_{e}$ experiments.

\begin{figure}[t!]
\begin{center}
\includegraphics*[bb=7 14 563 569, width=\linewidth]{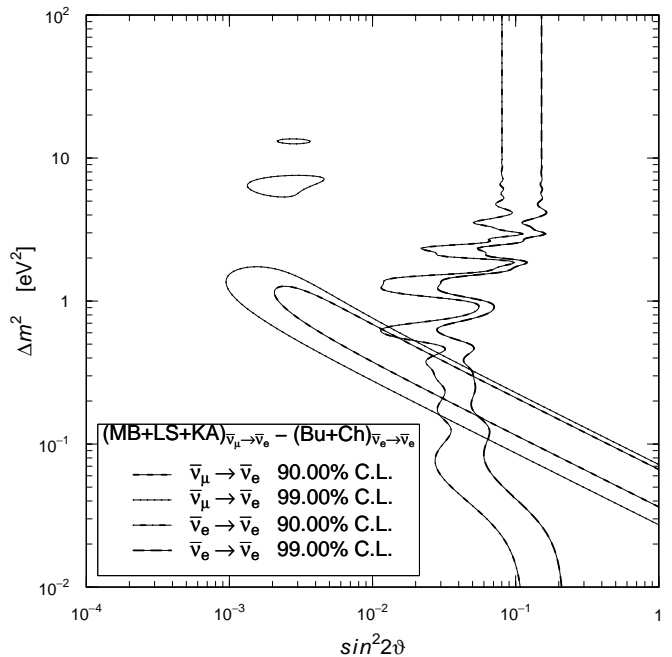}
\end{center}
\caption{ \label{031}
Superposition of the allowed regions in the
$\sin^{2}2\vartheta$--$\Delta{m}^{2}$ plane
obtained from the combined fit of MiniBooNE (MB), LSND (LS) and KARMEN (KA)
$\bar\nu_{\mu}\to\bar\nu_{e}$ data and
the exclusion curves obtained from the fit of reactor Bugey (Bu) and Chooz (Ch)
$\bar\nu_{e}\to\bar\nu_{e}$ data.
}
\end{figure}

Figure~\ref{031}
shows a superposition of the 90\% and 99\% C.L.
allowed regions in the
$\sin^{2}2\vartheta$--$\Delta{m}^{2}$ plane
obtained from the combined fit of MiniBooNE, LSND and KARMEN
$\bar\nu_{\mu}\to\bar\nu_{e}$ data and
the exclusion curves obtained in Ref.~\cite{0711.4222} from the fit of reactor
Bugey \cite{Declais:1995su} and Chooz \cite{hep-ex/0301017}
$\bar\nu_{e}\to\bar\nu_{e}$ data,
which currently provide the most stringent constraints on short-baseline reactor $\bar\nu_{e}$ disappearance.
The inequality (\ref{036})
implies that in Fig.~\ref{031}
the large-$\sin^22\vartheta$ part of
the straight region below $\Delta{m}^2\approx2\,\text{eV}^2$
allowed by the combined fit of MiniBooNE, LSND and KARMEN
$\bar\nu_{\mu}\to\bar\nu_{e}$ data
is excluded by
the results of reactor antineutrino experiments.
Quantitatively,
only the parts with
$\sin^22\vartheta \lesssim 3\times10^{-2}$
and
$\sin^22\vartheta \lesssim 5\times10^{-2}$
are allowed at
90\% and 99\% C.L., respectively.

\begin{figure}[t!]
\begin{center}
\includegraphics*[bb=5 11 571 571, width=\linewidth]{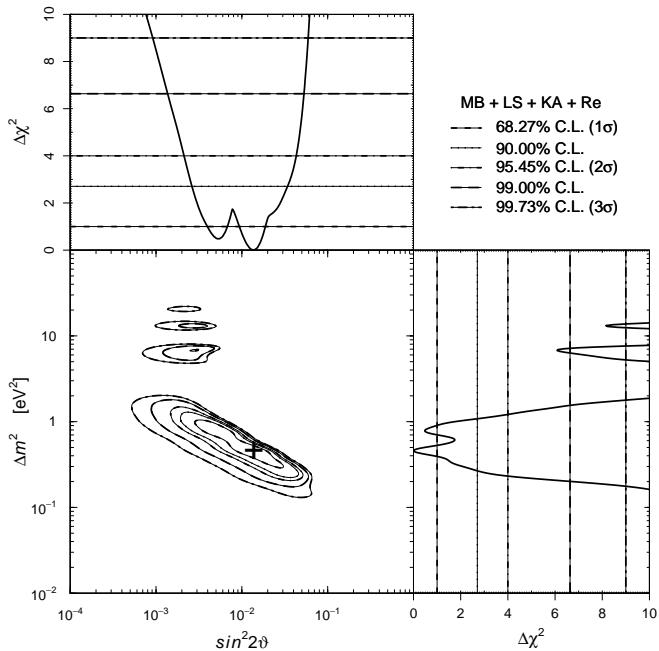}
\end{center}
\caption{ \label{032}
Allowed regions in the
$\sin^{2}2\vartheta$--$\Delta{m}^{2}$ plane
and
marginal $\Delta\chi^{2}$'s
for
$\sin^{2}2\vartheta$ and $\Delta{m}^{2}$
obtained from the combined fit of MiniBooNE (MB), LSND (LS) and KARMEN (KA)
$\bar\nu_{\mu}\to\bar\nu_{e}$ data and
the exclusion curves obtained from the fit of reactor Bugey and Chooz (Re)
$\bar\nu_{e}\to\bar\nu_{e}$ data.
The best-fit point is indicated by a cross.
}
\end{figure}

The inequality (\ref{036}) constrains the effective amplitude
$\sin^22\vartheta$
of short-baseline
$\bar\nu_{\mu}\to\bar\nu_{e}$
transitions, but does not allow a combined fit of
accelerator
$\bar\nu_{\mu}\to\bar\nu_{e}$ data
and
reactor
$\bar\nu_{e}\to\bar\nu_{e}$ data.
Such a combined fit can be done
if the upper limit in Eq.~(\ref{036}) applies,
i.e. if the inequality (\ref{036}) effectively becomes an equality.
This is the case if
$P_{\bar\nu_{\mu}\to\bar\nu_{e}} \gg P_{\bar\nu_{\alpha}\to\bar\nu_{e}}$
for
$\alpha \neq e, \mu$.
In the following we consider this interesting possibility,
which allows us to combine the accelerator and reactor data in order to find the
preferred region in the space of the oscillation parameters
which could be explored by future experiments
\cite{0909.0355,0910.2698,1007.3228,AndreRubbia:NEU2012}.

Figure~\ref{032} and the last column of Tab.~\ref{003}
give the results of the combined fit of
accelerator MiniBooNE, LSND and KARMEN
$\bar\nu_{\mu}\to\bar\nu_{e}$ data and
reactor Bugey and Chooz
$\bar\nu_{e}\to\bar\nu_{e}$ data
assuming an equality in Eq.~(\ref{036}).
The value of the parameter goodness-of-fit in Tab.~\ref{003}
shows that the accelerator and reactor data are compatible
under the hypothesis of $\bar\nu_{\mu}\to\bar\nu_{e}$ oscillations.

From Figure~\ref{032} one can see that
there is a favorite region at about 95\% C.L. around the best-fit point for
$2\times10^{-3} \lesssim \sin^22\vartheta \lesssim 5\times10^{-2}$
and
$0.2 \lesssim \Delta{m}^2 \lesssim 2 \, \text{eV}^2$.
Larger values of
$\Delta{m}^2$
are allowed only at more than about 95\% C.L.
for
$7\times10^{-4} \lesssim \sin^22\vartheta \lesssim 5\times10^{-3}$.

\section{\label{037}Conclusions}

We have considered the recent results of the MiniBooNE experiment \cite{1007.1150}
on short-baseline $\bar\nu_{\mu}\to\bar\nu_{e}$
oscillations,
which confirm the positive LSND signal \cite{hep-ex/0104049}.
Considering the simplest case of an effective two-neutrino-like
short-baseline oscillation probability
which depends on only two effective oscillation parameters,
$\sin^22\vartheta$
and
$\Delta{m}^2$,
we performed a combined fit of
MiniBooNE and LSND data
in order to find the allowed regions in the parameter space.

We considered also the results of the
KARMEN experiment
\cite{hep-ex/0203021},
in which $\bar\nu_{\mu}\to\bar\nu_{e}$ transitions have not been observed.
We have shown that the combined fit of
MiniBooNE, LSND and KARMEN data
is acceptable and leads to a shift of the region allowed by
MiniBooNE and LSND towards small values of
$\sin^22\vartheta$.

Finally, we have considered the model-independent bound on
short-baseline $\bar\nu_{\mu}\to\bar\nu_{e}$
implied by the limits on short-baseline $\bar\nu_{e}$ disappearance
obtained in reactor experiments.
From a combined fit of
accelerator
$\bar\nu_{\mu}\to\bar\nu_{e}$ data
and
reactor
$\bar\nu_{e}\to\bar\nu_{e}$ data
we have found that,
if the
$\bar\nu_{\mu}\to\bar\nu_{e}$
channel is dominant over other channels of flavor transitions
into $\bar\nu_{e}$,
the favored region of the effective oscillation parameters
lies within
$2\times10^{-3} \lesssim \sin^22\vartheta \lesssim 5\times10^{-2}$
and
$0.2 \lesssim \Delta{m}^2 \lesssim 2 \, \text{eV}^2$.
This region is interesting for a study of the possibilities to check the
LSND and MiniBooNE indication of
short-baseline
$\bar\nu_{\mu}\to\bar\nu_{e}$
oscillations
with future experiments
\cite{0909.0355,0910.2698,1007.3228,AndreRubbia:NEU2012}.

%

\end{document}